\documentclass[numbered]{trbunofficial}
\usepackage[table,x11names]{xcolor}
\usepackage{booktabs} 
\usepackage{siunitx}  
\usepackage{pdfpages}
\usepackage{subfig}
\usepackage{amsmath}
\usepackage{graphicx}
\usepackage[hidelinks]{hyperref}
\usepackage{ragged2e}
\justifying

\usepackage[hidelinks]{hyperref}
\usepackage[utf8]{inputenc}
\usepackage{pgf}
\usepackage{pgfplots}
\usepackage{tikz}
\usepackage{tikz-cd}
\usepackage{pgfplots}
\nolinenumbers
\AuthorHeaders{Akthar and Kucharski}
\title{Exploring computational complexity of ride-pooling problems}

\author{%
  \textbf{Usman Akthar}\\
  Faculty of Mathematics and Computer Science
  Jagiellonian University, Kraków, Poland\\
  usman.akthar@uj.edu.pl \\
  ORCiD: 0000-0003-4553-0550 
  \hfill\break \break
  \textbf{Rafał Kucharski}\\
  Faculty of Mathematics and Computer Science
  Jagiellonian University, Kraków, Poland\\
  rafal.kucharski@uj.edu.pl \\
  ORCiD: 0000-0002-9767-8883
  }

 \WordsPerTable{250}

 \TotalWords{5471}

\begin{document}
\maketitle

\section{Abstract}
\justify
Ride-pooling is computationally challenging. The number of feasible rides grows with the number of travelers and the degree (capacity of the vehicle to perform a pooled ride) and quickly explodes to the sizes making the problem not solvable analytically. In practice, heuristics are applied to limit the number of searches, e.g., maximal detour and delay, or (like we use in this study) attractive rides (for which detour and delay are at least compensated with the discount). 

Nevertheless, the challenge to solve the ride-pooling remains strongly sensitive to the problem settings. Here, we explore it in more detail and provide an experimental underpinning to this open research problem. We trace how the size of the search space and computation time needed to solve the ride-pooling problem grows with the increasing demand and greater discounts offered for pooling. We run over 100 practical experiments in Amsterdam with 10-minute batches of trip requests up to 3600 trips per hour and trace how challenging it is to propose the solution to the pooling problem with our ExMAS algorithm.

We observed strong, non-linear trends and identified the limits beyond which the problem exploded and our algorithm failed to compute. Notably, we found that the demand level (number of trip requests) is less critical than the discount. The search space grows exponentially and quickly reaches huge levels. However, beyond some level, the greater size of the ride-pooling problem does not translate into greater efficiency of pooling. Which opens the opportunity for further search space reductions.

\hfill\break%
\noindent\textit{Keywords}: Ride-pooling, Mobility as a service, Shared mobility, Complexity, Algorithms
\newpage

\section{Introduction}
\justify
Ride-pooling, offered by mobility platforms for travelers who agree to share their rides with co-travelers in exchange for reduced travel fare, is the promising but challenging emerging mode of urban mobility, especially with the rise of Mobility-on-Demand systems~\cite{bischoff2017city} such as Lyft\footnote{\url{http://www.lyft.me}}, Uber\footnote{\url{https://www.uber.com}}, car2go\footnote{\url{https://gotoglobal.com/en/}} and BlaBlaCar\footnote{\url{https://www.blablacar.pl}}. These mobility systems have the potential to mitigate congestion and increase accessibility via pooling~\cite{shaheen2019shared}. Shared mobility offers many benefits by reducing congestion and transportation costs~\cite{mourad2019survey}. Behind the overall potential offered by shared mobility, there are many challenges and open research problems.

Several studies have shown evidence that identifying attractive shared rides in ride-pooling services suffers from the \textit{curse of dimensionality} and is hardly solvable for large travel requests~\cite{engelhardt2020speed, alonso2017predictive, marivcic2021spatial,kucharski2020exact, liu2019bus}. The algorithmic challenge lies in the search space size, exploding combinatorically with the growing demand (number of travelers) and so-called willingness-to-share (behavioral or induced via pooling discounts). 
Therefore, ride-pooling is typically solved by heuristics or for small instances~\cite{simonetto2019real}. 
The problem of combining trips into attractive shared rides and their willingness-to-share for real-size demand level remains challenging. As argued by Santi et al.~\cite{santi2014quantifying} computing the pairwise shareability graph is already challenging and the complexity adding extra degrees (number of travllers) will limit the potential of ride-pooling in practice. 

Many efforts have been made to circumvent the problem of identifying the attractive shared rides on realistic datasets. Most notably: Alonso-Mora et al.,2017~\cite{alonso2017demand}, Santi et al.~\cite{santi2014quantifying} and de Ruijter et al. 2020~\cite{de2020ride}, they either (a) only consider the limited number of shared ride requests within the same time-window, (b) only consider the search space of lower degree with less combinatorial shared trips without performing efficient graph search, (c) only relying on the heuristics which limit the potential of ride-sharing in real-world scenarios. While most ride-pooling solutions~\cite{danassis2022putting} apply a real-time approach when requests are not known in advance but appear in real-time. In practice, requests are often 'batched' in the fixed interval (e.g., 5 or 10 minutes) and pooled later for efficiency. Obviously, with the longer batches, the solution may become more efficient (though more computationally demanding), yet at the cost of lower flexibility for the users (travelers).

Most common approaches~\cite{alonso2017demand, santi2014quantifying} offer fixed-windows approach, where any pooling sequence with delay and detour below a predefined maximal is allowed and equally attractive for all the co-travellers. On the contrary, we used utility-based filtering \textit{best fit} for attractive rides. 

\begin{figure}[t]
    \centering
    {\includegraphics[clip,  trim=0.8cm 0.3cm 1.5cm 0.7cm, scale=0.75]{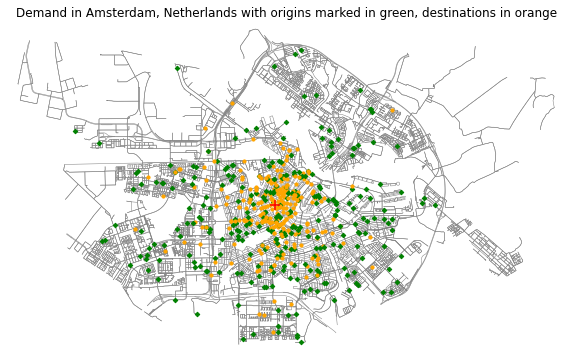} }

        \caption{\textbf{Synthetic demand for ride-pooling in Amsterdam, Netherlands used in our experiments. The origins (green) are more dispersed then destinations (orange) which concentrate more around the city center (red cross) to reproduce the structure typical to the morning peak hour. In our experiments we keep the same spatial distributions and vary the number of trips.}}
        \label{fig:DemandAmsterdam1}%
\end{figure}



In our paper, we seek to unpack: the computational complexity of real-world ride-pooling problems and identify how the complexity grows in various settings for the case study of ride-pooling in Amsterdam, The Netherlands.  We traced the computation times, memory consumption, search space sizes, and properties of underlying shareability graphs. We trace it against two control variables: a) demand levels, i.e., number of trip requests in time, and b) the discount offered to attract travelers to ride-pooling.  Thanks to such an analysis, we may better understand if and how the ride-pooling problems may become solvable at acceptable times for growing problems. 

In particular, we ask whether the growth is linear or faster. Which properties of the ride-pooling problem have the highest significance for problem complexity. Does the complexity have equally impact: time, size, and memory consumption? Can we relate the complexity of the problem to the efficiency of ride-pooling? These are open research problems that we address with this experimental study.

We identified that beyond some level of demand and discount, the significant increase in search space does not translate into a more efficient pooling solution. Although the theoretical sizes of search space quickly become intractable, utility-based approaches can still exhaust it. Yet, only to some extent: Under high demand levels, discounts, and willingness to share this problem explode. We identified this limit, at the level of 40\% discount and 400 trip requests per 10-minute batch.

Overall, the main contribution of this paper lays in observing and reporting the computational complexity of ride-pooling with the goal to: (a) analyze the search space size exploitability for ride-pooling at large-scale settings, and understand (b) what are the limits under which ride-pooling remains solveable.

The rest of the paper is organized as follows. Section 2 introduces the proposed methodology for characterizing the complexity of ride-sharing. Section 3 presents the experimental results and the analysis. Section 4 concludes the paper.  




\section{Methodology}


Our methodology revolves around the ExMAS algorithm proposed by Kucharski et al.,2020~\cite{kucharski2020exact}, an offline algorithm that addresses the complexity of the ride-sharing problem via the utility-driven approach. Below, we first formulate theoretical computational complexity of ride-pooling, then we move to a brief introduction of the idea behind ExMAS algorithm and its approach to ride-pooling. Finally we report measures that we trace while answering the questions of ride-pooling complexity.


\begin{figure}[!b]
    \centering
    
    
    {\includegraphics[clip,  trim=0.1cm 0.1cm 0.1cm 0.1cm, scale=0.7]{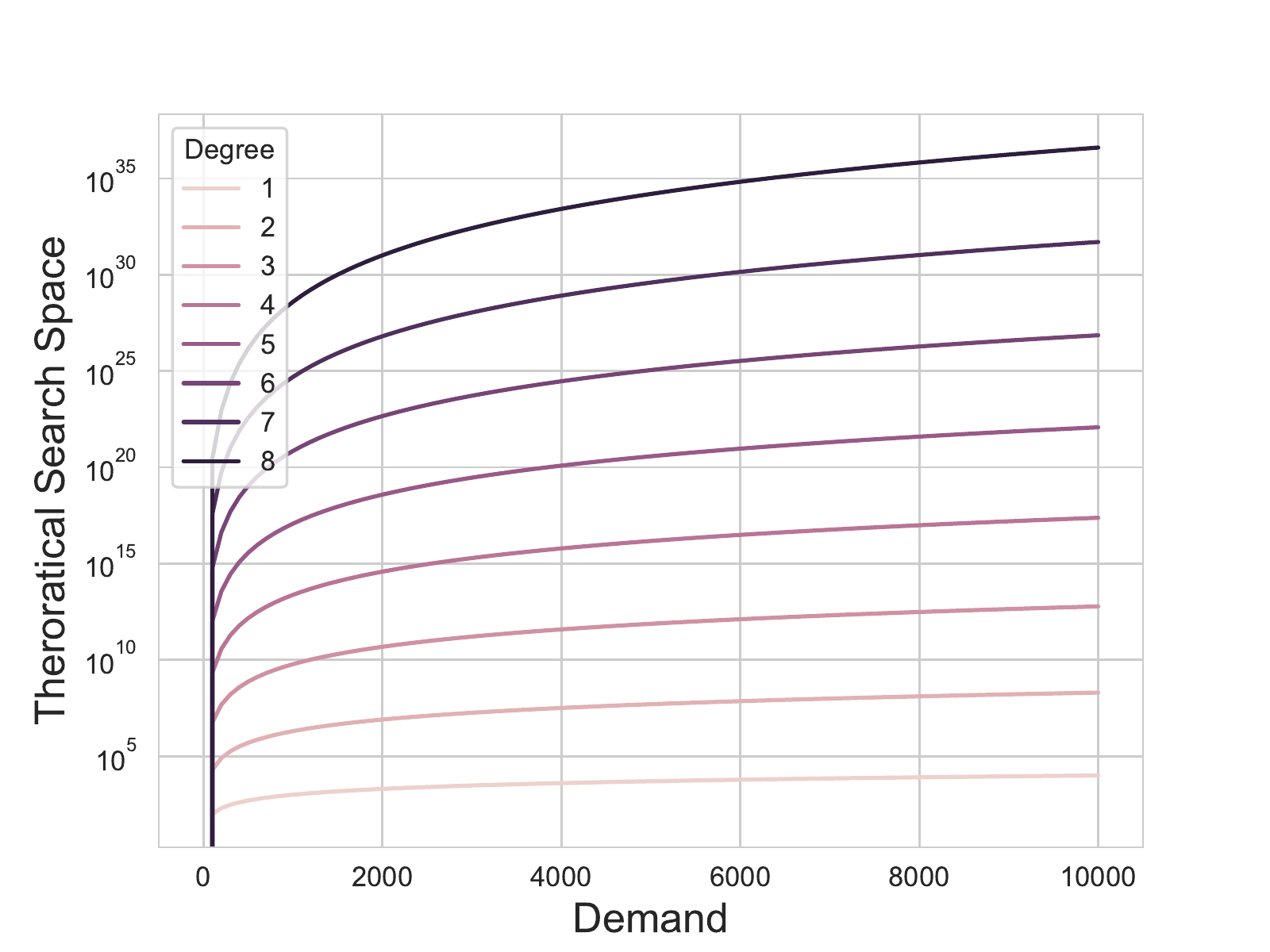} }%
        \caption{\textbf{Theoretically computed search space of ride-pooling problems. Lines denote how many feasible combinations of shared rides can be theoretically constructed for a given demand size (number of trips on x-axis) and when pooled rides size (degree) changes (which is denoted with a different line colours.)}}
        \label{fig:TheoraticalSpace}%
\end{figure}

\subsection{Theoretical Search Space for Ride-Pooling Problem}

In the ride-pooling we aim to identify feasible combinations of travelers of a given degree (number of co-travelers). The complexity of such search grows with the number of travelers and increasing degree, which are key variables of the ride-pooling problem complexity.
For the $n$ trip requests, there are $n^{2}$ potential pairs that could share a ride, each of them with four possible orders of pickups and drop-off of two co-travellers. This already big number quickly grows for increasing degrees, i.e., when more travellers potentially sharing rides are considered. In equation~\ref{searchspace}, we propose theoretical formula for search space complexity without any filtering and heuristics. The search space $S$ is calculated as a number of possible subsets of size $d$ in the set of all the travelers requesting a pooled rides $Q$. This is further multiplied with the order in which these travelers are picked up ($d!$ combinations), and dropped-off ($d!$ again), which yields a theoretical formula of:

\begin{equation} \label{searchspace}
	S =  \binom{Q}{d} d! d!
	\end{equation}

In Figure~\ref{fig:TheoraticalSpace}, we illustrate the theoretically computed search space for the ride-sharing problem. The search space grows significantly in the log-scale and the growth is huge with respect to the number of travelers requesting pooled trips (x-axis), yet more importantly to the number of travelers riding together (degree). The latter is critical to make the ride-pooling sustainable and actually reduce the congestion via the occupancy levels comparable to public transport. To find an optimal  ride-pooling solution, a large computation is required, and the search space explodes with $10^{35}$ size, making it impossible to search exhaustively. In particular, for 2000 travelers there are $10^{32}$ feasible rides of $8^{th}$ degree, while for 8000 travelers there are $10^{15}$ feasible rides of $4^{th}$ degree.


\subsection{On Utility-Based Filtering for Attractive Rides with ExMAS}

To overcome the curse of dimensionality for real-size demand patterns, the utility-driven search space method is applied to effectively explore only attractive shared rides and avoid unnecessary searches for acceptable computation time~\cite{kucharski2020exact}. Here we briefly introduce the main concept, which is detailed in the full ExMAS paper and available at the public repository\footnote{\url{https://github.com/RafalKucharskiPK/ExMAS}}. 
ExMAS algorithm utilize the utility based sequential reduction to efficiently reduce the curse of dimensionality and make the computation tractable in practice. Otherwise, resulting in memory and/or running time issues.


More specifically, ExMAS was developed for offline matching to improve filtering for attractive rides based on utility-based approach.  The rides of second-degree are identified based on pairwise searches. We have adopted the utility-based function, considering shared rides are attractive if ride-pooling utility is greater than nonshared alternatives~\cite{kucharski2020exact}. Leveraging on the utility-based filtering, we formulate the difference between the shared and non-shared rides and use this formula to filter for attractive pooling only:

\begin{center}
    \begin{equation} \label{eqn}
	\Delta U = U^{s}-U^{ns}=\beta^{c}\lambda l+\beta^{t}\left ( t- \beta^{s} \left ( t^{s} + \beta^{d}t^{d} \right )\right ) 
	\end{equation}
\end{center}

In equation~\ref{eqn}, we compute attractiveness $\Delta U$ of pooling with $U^{s}$ and $U^{ns}$ a difference between the pooling and private ride utilities. The direct travel time $t$ of a private ride is now detoured due to pooling $t^s$ and possibly delayed $t^d$. We also consider that $\lambda$ stands for the discount for sharing a ride to compensate for the downside of shared rides in terms of longer travel time. We explicitly consider only trips with $\Delta U>0$.

Other parameters that affect the search space are $\beta^{c}$, $\beta^{t}$ and $\beta^{d}$ for cost sensitivity, sharing discomfort, and delay sensitivity respectively. However, we did not experiment with them in this study and resort to the only one directly controllable by the operator: the discount. The key control of the demand size is the discount rate ($\lambda$) that is offered for the shared rides, the optimal value of the $\lambda$ encourage both by the service provider and the traveler.

\subsection{Performance Indicators and Scalability for Large-Scale Demand}

The goal of our methodology is to analyze the impact of computation time to enumerate all feasible attractive rides in the case of Amsterdam, The Netherlands. 
We proposed the performance indicators to measure the complexity of the search space to understand if the solution employed (i.e. ExMAS) solves the practical ride-pooling problems in a reasonable time frame (running time). We observe the complexity via the following indicators:
 %

\raggedright
\begin{itemize}
    \item search space size, i.e. number of explored and queries feasible ride candidates identified and processed with ExMAS algorithm (which - thanks to an effective utility-driven filtering - is merely a fraction of search space computed theoretically with eq. 1).
    \item computation time, i.e. time to solve the ride-pooling problem, from pairwise shareable rides, up to rides of greater degree and finally the solution of the bipartite matching problem.
    \item shareability graph attributes, i.e. topology of the underlying graph constructed to solve the ride-pooling problem. We trace this parameter to understand how the complexity relates to graph properties.
    \item ride-pooling efficiency, which we use to understand if the problem size determines how effective the ride-pooling is. We trace it with the two attributes: a) ratio of the pooled rides in the solution - i.e. how many travelers share rides, b) utility improvements (relative improvement of utility gained by travellers due to pooling) - i.e. total of $\Delta U$ from eq. 2 for all the travelers.
\end{itemize}



\section{Results}
\justify
This section is devoted to present the results of our evaluation. We performed an evaluation on mid-size city Amsterdam in Netherlands (below 1 million inhabitants). The goal of our experiment is to examine the number of indicators for computational complexity across the experimental scheme. These results may provide the foundation for the development of efficient methods and heuristics for solving the practical ride-pooling problems. 

We first report running times and search space resulting from the experiments in Figure~\ref{fig:RuntimeVSearchSpacve}. Then we detail the search space for this rides of second and third degree in Figure~\ref{fig:Degree2VSDegree3}. We then move to show the pooling efficiency obtained for the experiments in Figure~\ref{fig:SearchSpaceDegree} and conclude with showing how it relates to the search space in Figure~\ref{fig:AverageNode} and in Figure~\ref{fig:Poolingefficiency}.



\subsection{Experimental setup}
We run the ExMAS algorithm for the detailed network of Amsterdam (downloaded with \texttt{osmnx}\cite{boeing2017osmnx}) with a synthetic demand of varying demand levels. We run experiments across the 2-dimensional grid of parameters: demand levels (trip requests per hour) and discount offered for pooling (relative trip fare reduction related to the private ride). 
In both dimensions we intended to reach the limits of tractability, i.e. when search space would become too big for the algorithm to compute. In practice, we have reached it only for the discount, which happened to be more critical for computations. 
We experimented with up to 3,600 requests for trips per hour (600 per batch of 10 minutes), discounts of up to 40\%.  More specifically, we explored the values of the shared discount range of $\lambda \approx \left \{ 0.05, 0.10, 0.2, 0.25, 0.3, 0.35, 0.40 \right \}$ and demand levels (number of trips requested in a 10-minute batch) ranging from 50 to 600 (sampled with increment of 50 travellers).  In the case of the Amsterdam demand, we use a synthetic demand generation procedure, where origins and destinations are placed at random nodes (of the detailed Amsterdam network) with a given probability depending on the distance from the center of the city. We used plausible ride-pooling setting with actual values-of-time, travel times and ride-hailing pricing. We focus on shared-rides of degree up to four passengers in the experiments. 
In each experiment, we use the same spatial distributions but vary the number of trips so that demand generation is non-deterministic and a single demand set is just one realization.

We conducted experiments on the 72-core, 2x Intel Xeon E5-2697 v4 server with 128GB RAM. We traced: computation times, search space, memory consumption, ride-pooling efficiency, and topology of underlying shareability graph (see details in \cite{kucharski2020exact}). We recorded them at the initialization, at each processed degree of pooled rides (first when the pairs were explored, then triples, quadruples, until the maximal degree was explored) and when the so-called matching (the matching problem on the bipartite graph) was conducted. Our evaluation is a strategic-offline, i.e., we pool the batch of up to seven hundred trips requested in 10 minutes. Such an approach resembles both the practical real-time problem and allows us to reach the significant complexity of the pooling problem.

\begin{figure}[!b]
    \centering
    \hspace*{-0.7cm} 
    \subfloat[\centering Number of feasible rides explored with ExMAS algorithm (search-space) increases with the demand levels (x-axis) and the discount offered (line colours). While for the batch of 500 trips per 10-minute and 15\% discount the search space of ca. 10000 feasible rides needs to be explored, when the discount is increased to 30\%, search space grows to $10^5$ rides. The critical point is reached at the 40\% discount and 400 trip requests- then the search space has size above $10^5$ and ExMAS failed to compute.]{{\includegraphics[clip,  trim=0.2cm 0.1cm 1.5cm 0.7cm, width=8cm]{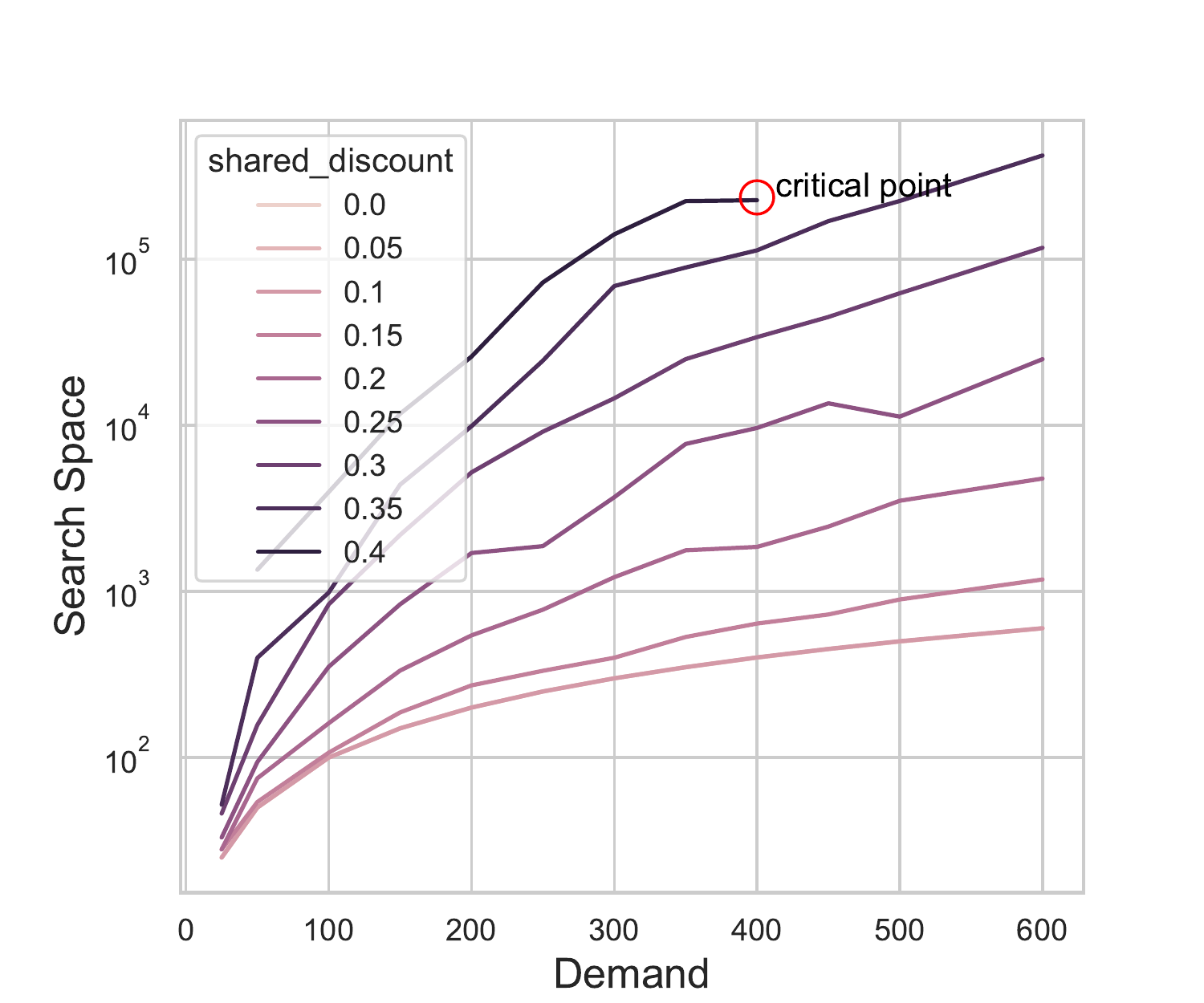} }}%
    \qquad
    \subfloat[\centering Running time needed to solve ride-pooling problems with ExMAS. It grows significantly with the demand size (x-axis), yet much sharper growth is visible with increasing the discount (line colours). While for 20\% discount 200 requests takes ca 10s to compute, for 500 trip requests it grows to 100 seconds. Yet for 600 trip requests this 100 seconds grows to 2.8 hours when discount is increased - which is hardly acceptable for real-time ride-pooling problems.]{{\includegraphics[clip,  trim=0.2cm 0.1cm 1.5cm 0.7cm, width=8cm]{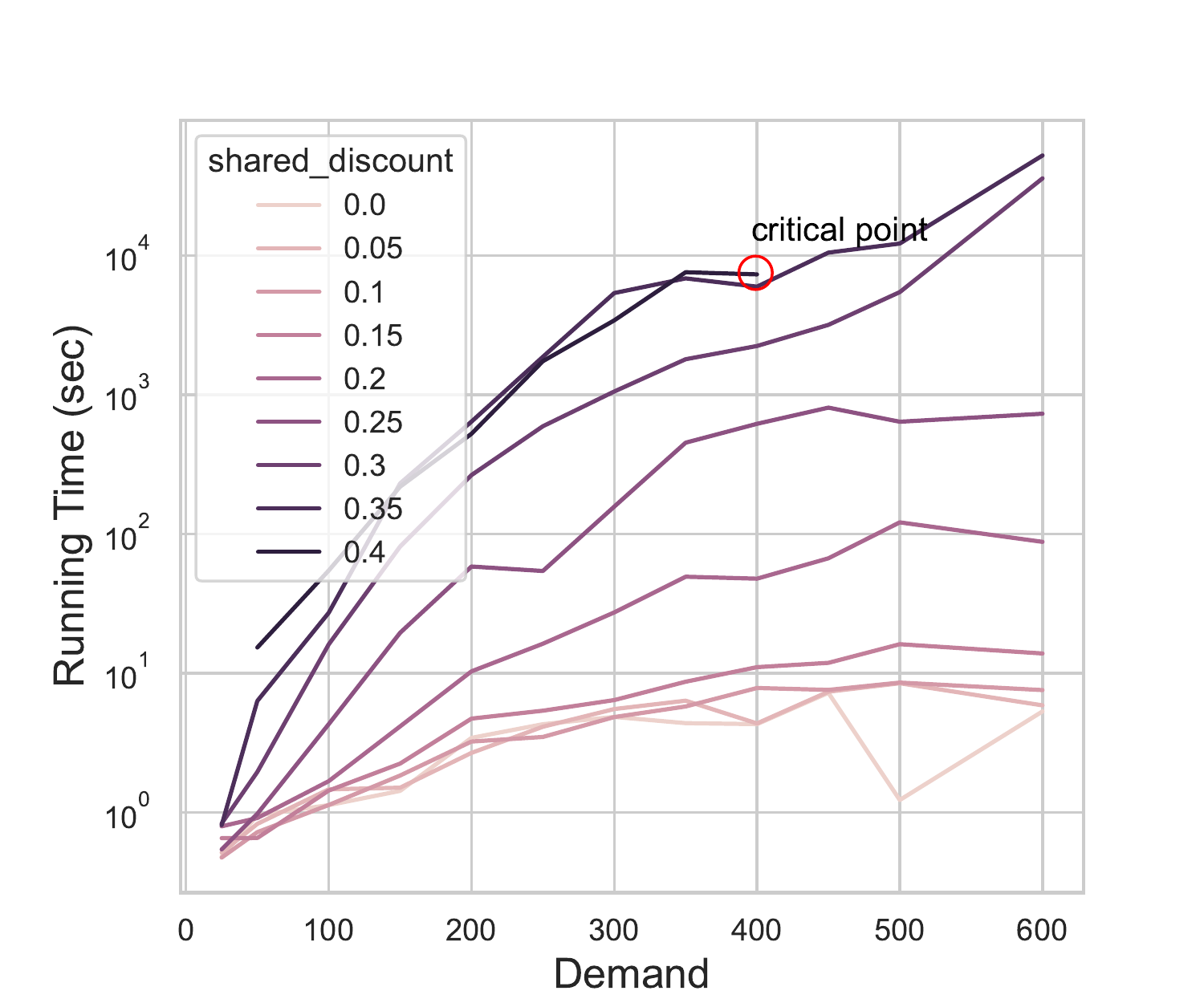} }}%
    \caption{\textbf{Showing the impact of  $\lambda$ on running time and search space complexity. By varying the demand and discount effects the overall performance, until reaches a critical point; where computation become intractable.}}%
    \label{fig:RuntimeVSearchSpacve}%
\end{figure}

When the ExMAS algorithm reaches its critical point, as shown in Figure~\ref{fig:RuntimeVSearchSpacve}, we identified meaningful flags to stop the algorithm from exploding, namely, by exploring the attributes of the underlying shareability graph. We plan to follow up on this and adjust the algorithm to extend its capabilities. We stored the output of each of the algorithm runs in the csv files for further analysis.

\subsection{Ride-pooling problem complexity}
In the following, we present an analysis of how the shared discount affects the size and running time of the search space.  We can see that both the demand and discounts for shared rides $\lambda$ have a strong impact on the size of the search space. 
The problem becomes intractable even for 400 travelers, with a 40\% discount. The search space grows non-linear with respect to all attributes, and we found a strong correlation with all input attributes and the search space size explodes as shown in Figure~\ref{fig:RuntimeVSearchSpacve}(b). 

The shared discount has the greatest impact on the running time and search space, until it reaches a critical point and from the computational complexity perspective it becomes intractable, as shown in Figure~\ref{fig:RuntimeVSearchSpacve}(a). 
Mind that despite the strong trends observed also for the computational times, there are some discrepancies presumably due to asynchronous, job-based multithreaded server, where some processes may have terminated faster due to higher priority in the job queue or more available resources (as observer e.g. for discount of 40\% which had more resources at the end of the queue and was calulated faster despite the larger search space).

\subsection{Trips of higher degree and complexity}
\begin{figure}[!b]
    \centering
    \hspace*{-0.7cm} 
    \subfloat[\centering Number of feasible rides of second degree identified for various demand levels (x-axis) and discounts (colours).]{{\includegraphics[clip,  trim=0.2cm 0.1cm 1.5cm 0.7cm, width=8cm]{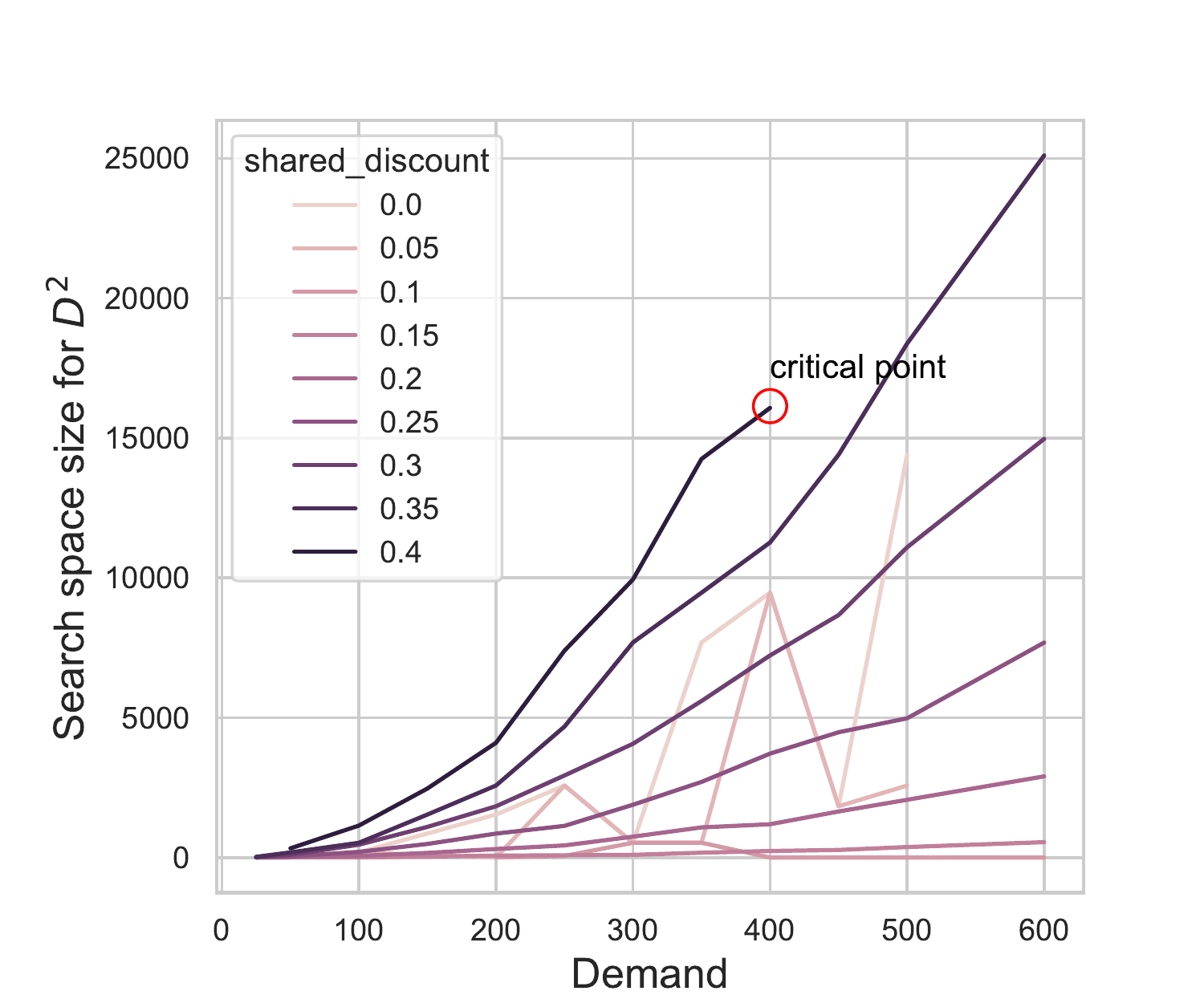} }}%
    \qquad
    \subfloat[\centering Number of feasible rides of third degree (shared by three co-travellers).]{{\includegraphics[clip,  trim=0cm 0.1cm 1.5cm 0.7cm, width=8cm]{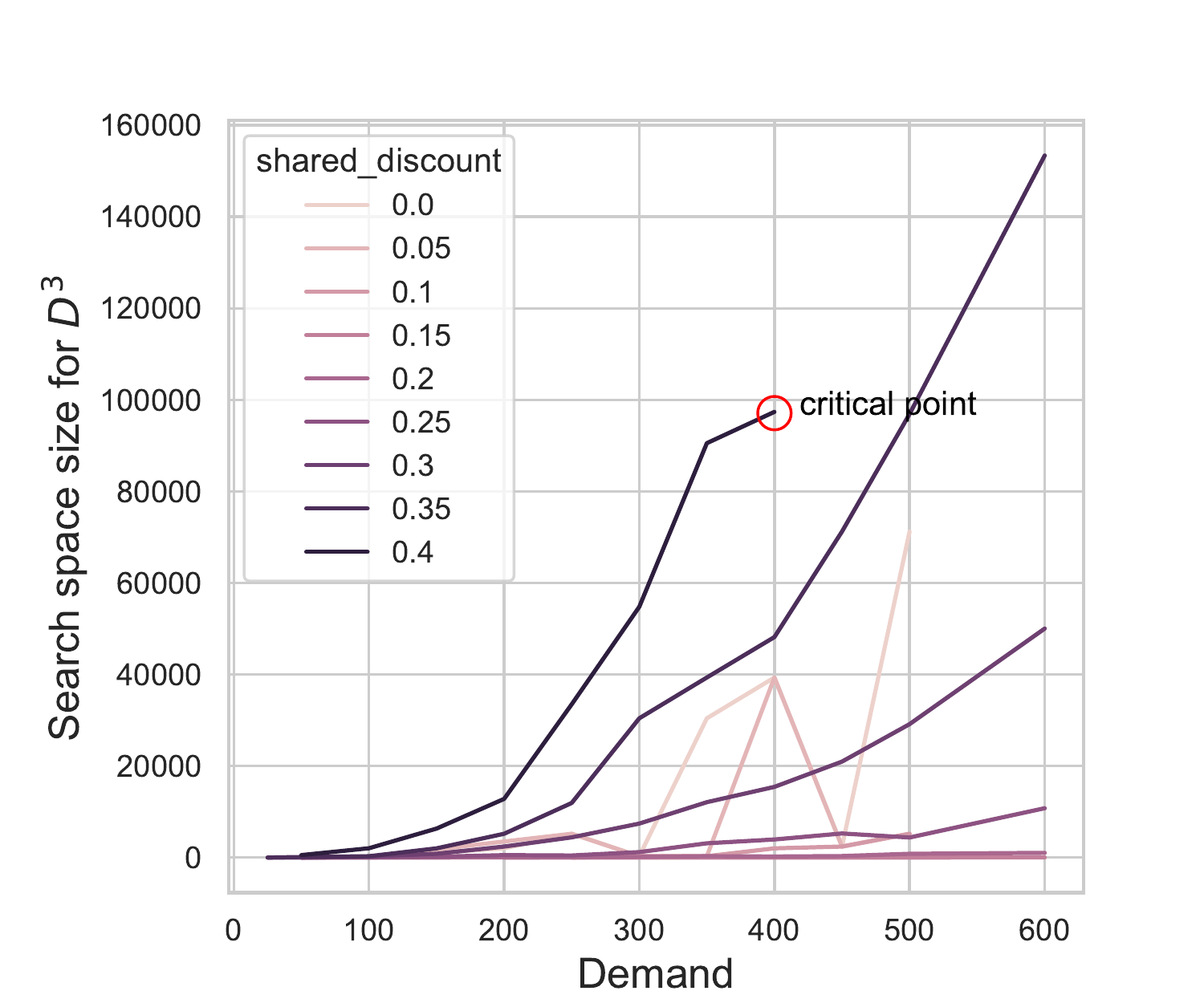} }}%
    \caption{\textbf{Number of feasible rides of second degree (a) grows with the demand size. For the lower discount levels the relation is linear, yet when greater discounts are offered, number of identified feasible pairs grows exponentially. This trend is even more significant for the rides of third degree. For low demand and low discounts ExMAS does not identify rides shared by three travelers. Yet for the batch of 600 travellers 15000 triples are identified when pooling is discounted with 35\%.}}%
    \label{fig:Degree2VSDegree3}%
\end{figure}

The size of the search space, like in theoretical examples, grows significantly when more than two travellers can share a ride. As for the lower discount levels, the trends seem to grow linearly and then exponentially grows for greater discounts. In our examples this was visible only when sufficient critical mass was reached, as shown in Figure\ref{fig:Degree2VSDegree3}. Mind that number of explored rides shared by two travellers does not exceeded 25 000, while the number of triples reached 150 000 trips. Which, has presumably more significant impact on computation times as shown for the case of rides shared by three travelers on Figure~\ref{fig:Degree2VSDegree3}(b). The non-smooth observations of trends for lower discounts show high variability of ride pooling under some settings and call for more replications to reproduce the statistically sound distributions in future studies.

\subsection{Efficiency and Computational Complexity}

\begin{figure}[!t]
    \centering
    \hspace*{-0.7cm} 
    \subfloat[\centering Efficiency of pooling rides (reduction in the perceived costs - i.e. utility as formulated in eq. 1) the trend is most visible with the discounts offered, and (to lesser extent) with the demand levels. The critical mass of ca. 300 travelers in the 10-minute batch is needed to stabilize the benefits of pooling. ]{{\includegraphics[clip,  trim=0.1cm 0cm 1.5cm 0.7cm, width=8cm]{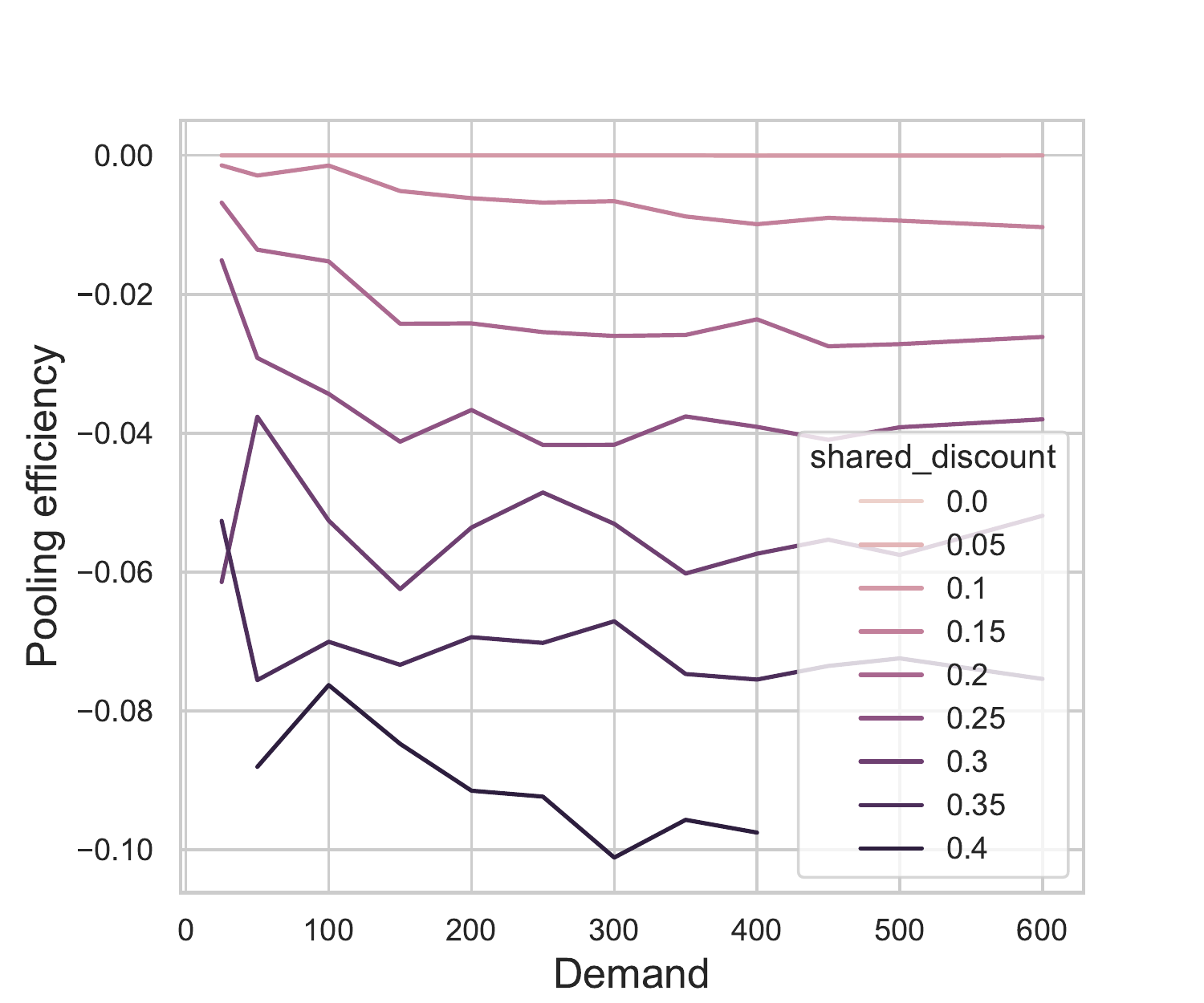} }}%
    \qquad
    \subfloat[\centering Share of rides that are actually pooled. It grows both with the demand level as well as the discounts offered. It exceeds 80\% for discounts greater than 25\% and demand levels above 300 - beyond which it stabilizes]{{\includegraphics[clip,  trim=0.1cm 0cm 1.5cm 0.7cm, width=8cm]{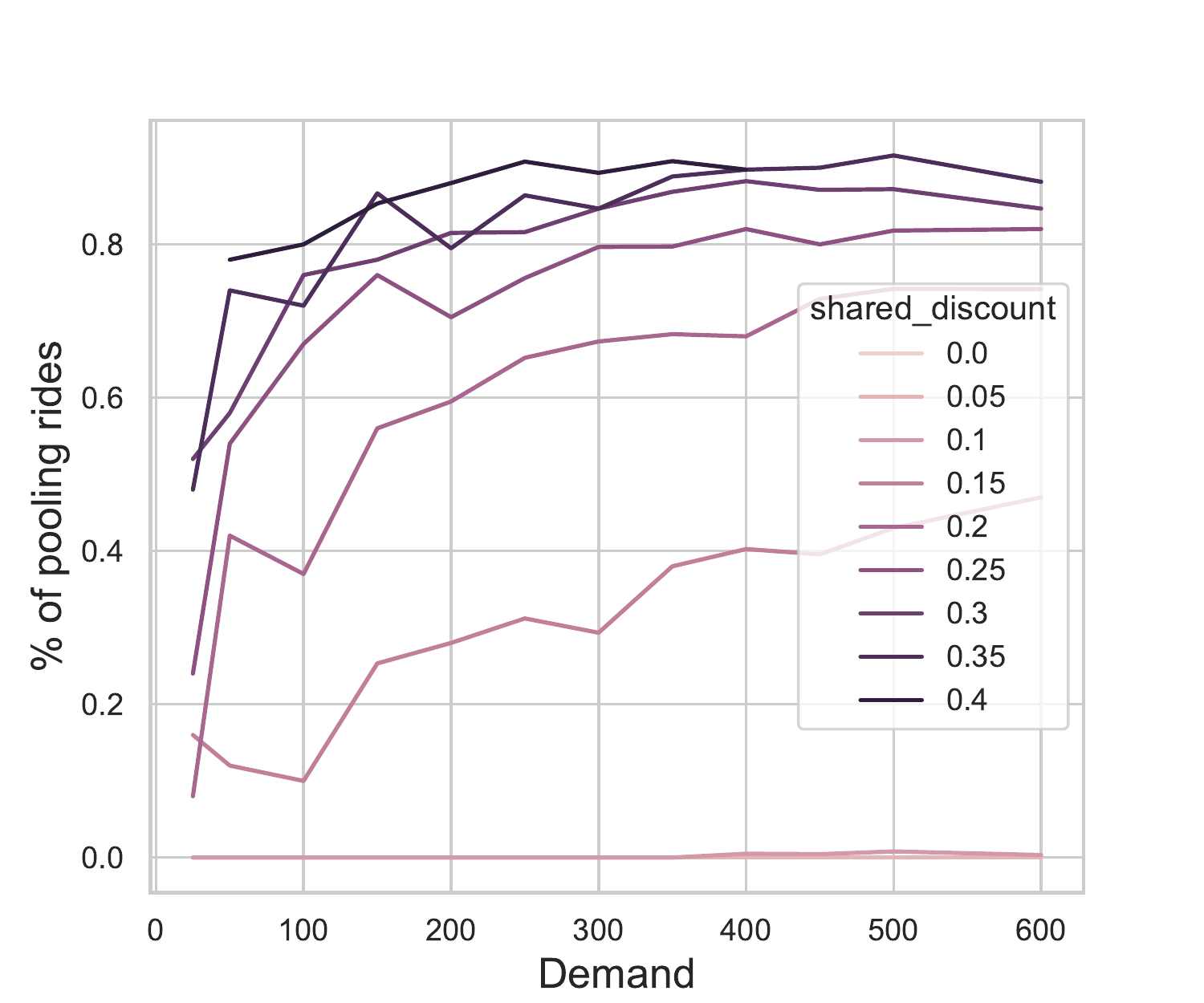} }}%
    \caption{\textbf{ The effect of discounts on the efficiency of ride sharing in various experimental environments. }}%
    \label{fig:SearchSpaceDegree}%
\end{figure}

We presumed that greater search space of the pooling problem would translate into more efficient system, where more compatible pooling groups may be identified. This is not so obvious, as we report below.
We analyze the impact of sharability on ride-pooling efficiency with different discounts. The trend becomes more obvious when offered a higher discount, as shown in Figure~\ref{fig:SearchSpaceDegree}(a). We plot it against two variables: demand level (i.e., number of trips) and max degree level of pooled ride (number of travelers that can share a ride together).  We can see that the percentage of pooled rides grows with the demand level and eventually increases the search space. Furthermore, the increasing percentage of combined rides becomes evident when offering a higher discount, as depicted in Figure~\ref{fig:SearchSpaceDegree}(b).

\begin{figure}[!t]
    \centering
    \hspace*{-0.7cm} 
    \subfloat[\centering Relation between pooling efficiency (y-axis) and search space size. While there is a strong relation it is not always linear and evident. For instance the benefits of pooling are stable at the 7\% for search space sizes varying from $10^3$ to $10^6$]{{\includegraphics[clip,  trim=0cm 0.1cm 0.1cm 0.1cm, width=8cm]{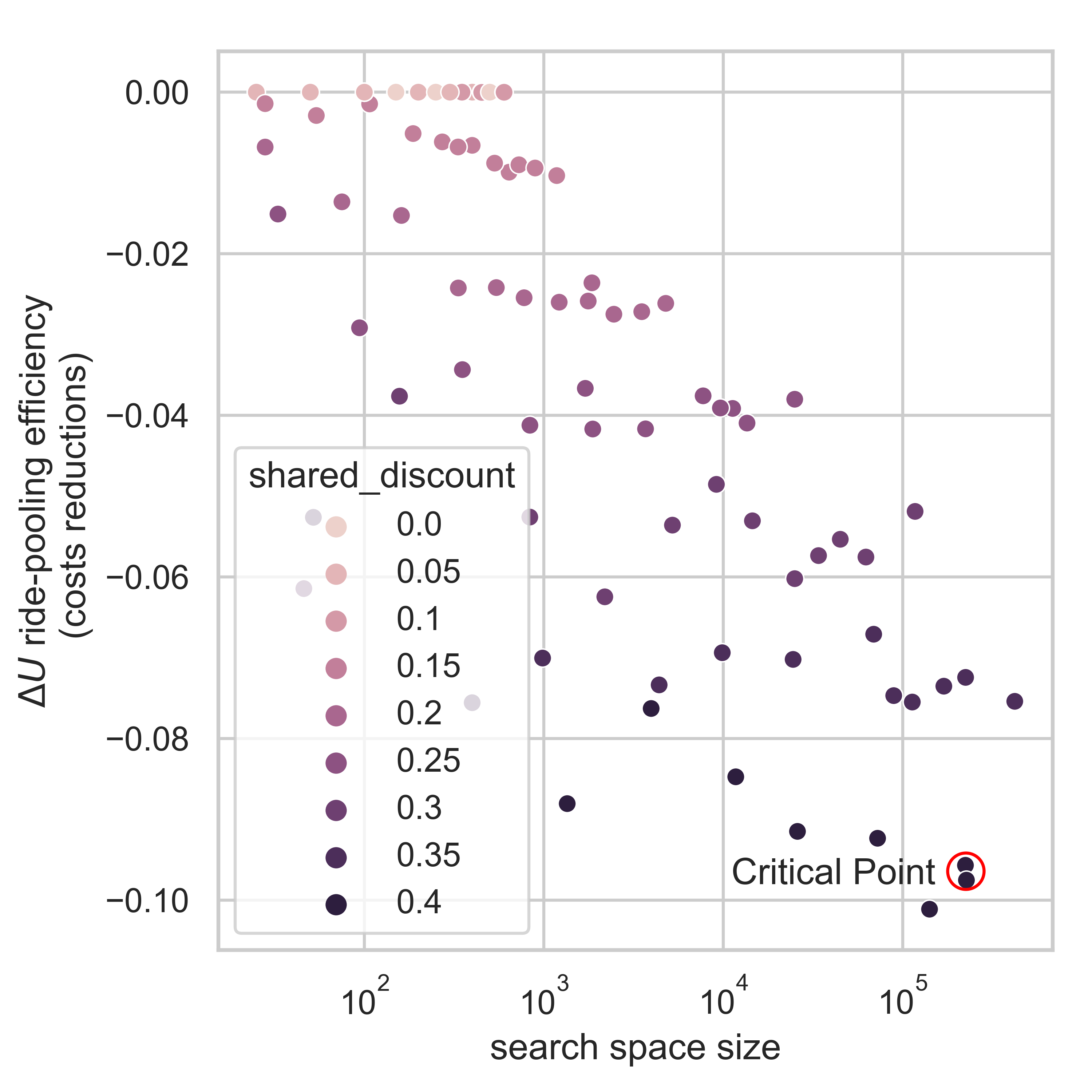} }}%
    \qquad
    \subfloat[\centering Strong trend (on log-log plot) between the properties of the so-called pair-wise shareability graph (x-axis) and search space of the problem. While for the lower demands and discounts there is still a significant dispersion, the trend becomes more profound with increasing search space. ]{{\includegraphics[clip,  trim=0cm 0.1cm 0.1cm 0.1cm, width=8cm]{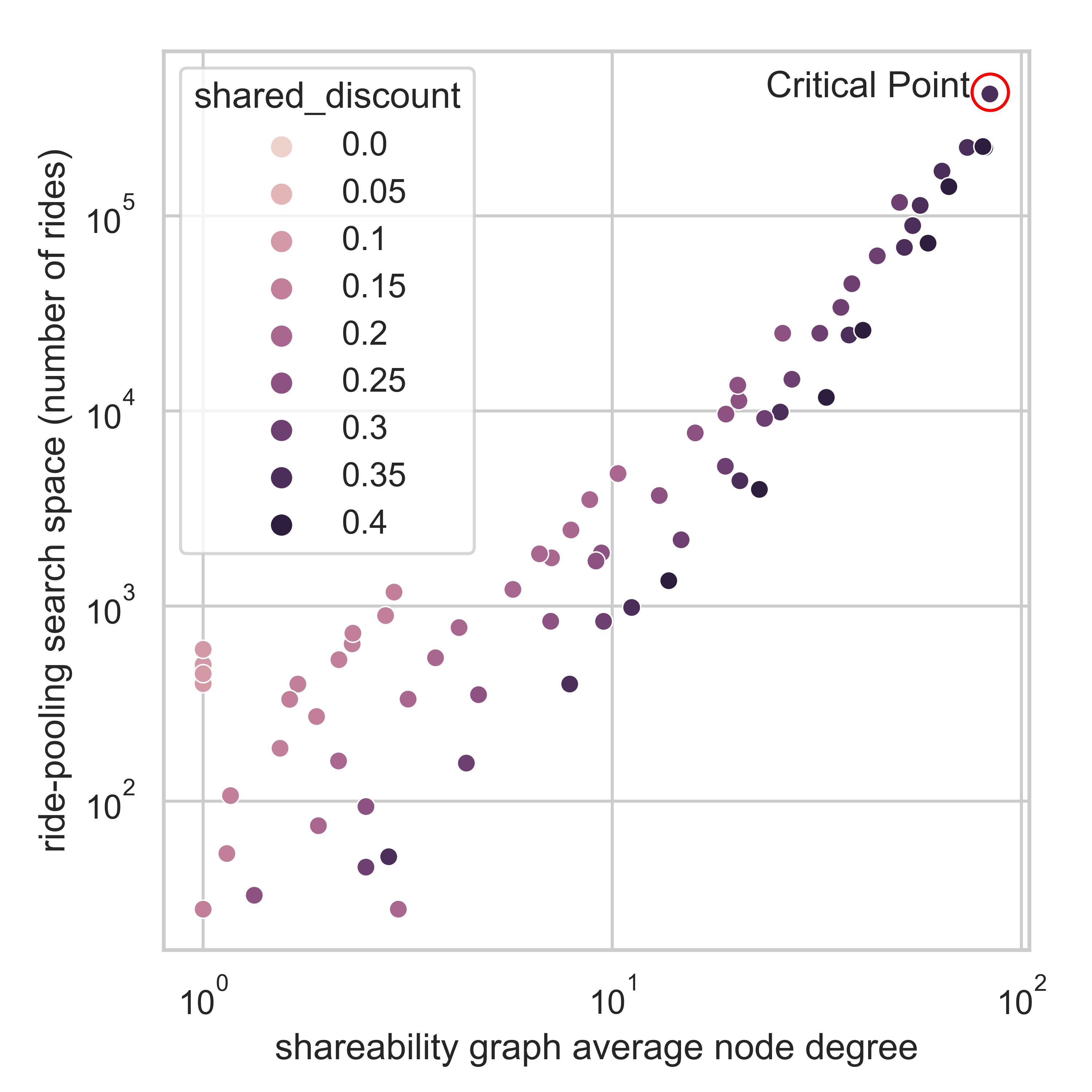}}}%
    \caption{\textbf{ Showing the pooling efficiency of large-scale search space and gains from trip sharing expressed  in term of its shareability graph, a strong trend between the shareability graphs and search space.}}%
    \label{fig:Poolingefficiency}%
\end{figure}

We see that efficiency of ride-pooling plateaus beyond some level, yet the search space grows, as shown in Figure~\ref{fig:Poolingefficiency}.
We observe a strong trend for the cost reduction of the efficiency of pooling $\Delta U$, as shown in Figure~\ref{fig:Poolingefficiency}(a), with a minimum of less than 10\%, we notice a less willingness of the traveler to share a ride. But a higher discount price with less profitably explodes the search space, and finding the optimal search pairs requires computational cost until it reaches its critical point. For smaller search spaces of ride-pooling problems, the shareability graph is sparse, so the number of attractive rides increases with degree. But for the higher discounts, the graph becomes too dense and cannot be effectively traced in reasonable time, as shown in Figure~\ref{fig:Poolingefficiency}(b). Since the complete search space is intractable, researchers need to resort to heuristics.

\begin{figure}[!t]
    \centering
    \hspace*{-0.7cm} 
    \includegraphics[clip,  trim=0.2cm 0.1cm 1.5cm 0.7cm, width=8cm]{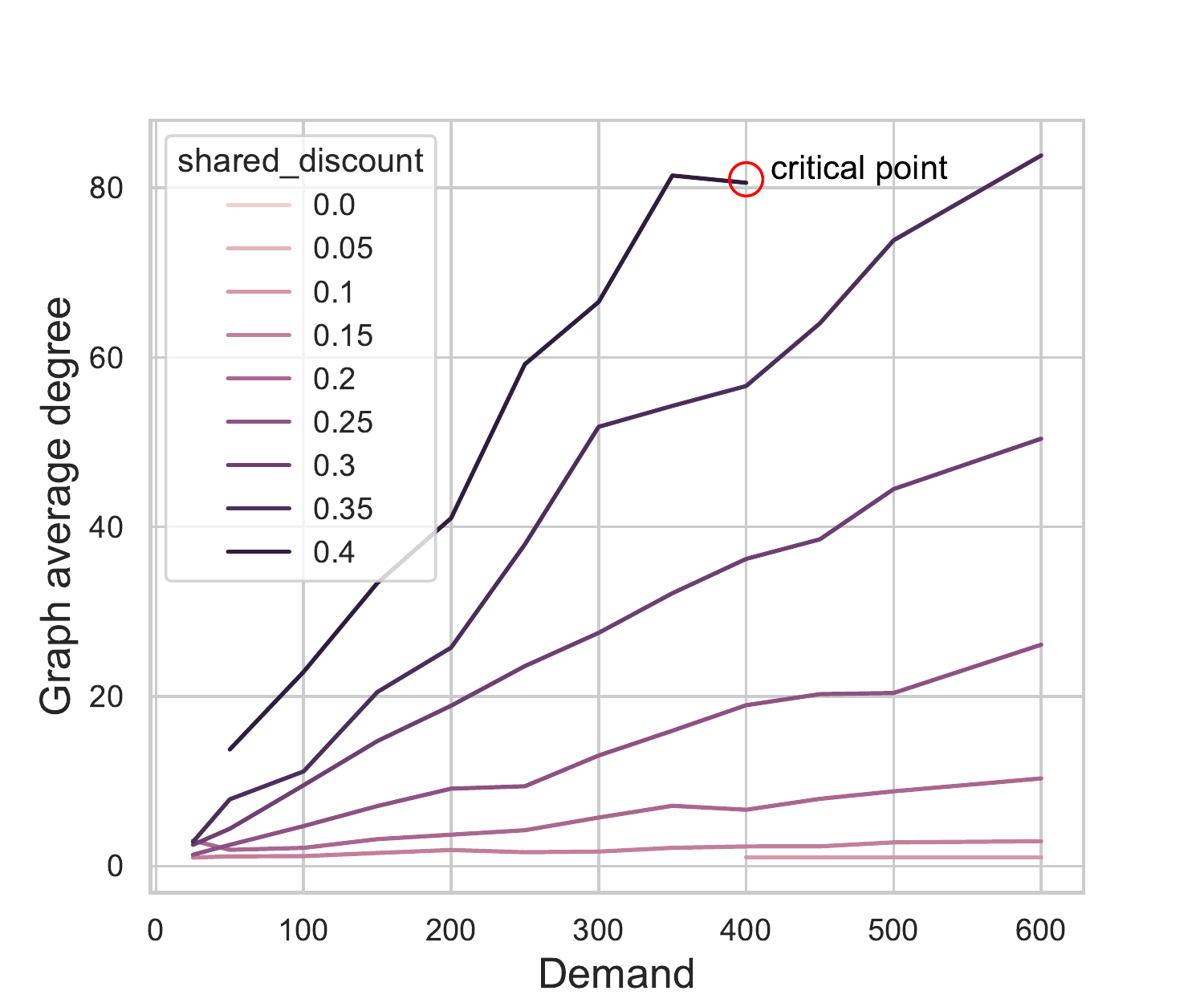} 
    \caption{\textbf{Average node degree of the so-called shareability graph (detailed in \cite{kucharski2020exact}) growing with demand levels and discount. For big sizes of demand problems each traveler can share a ride with average of 80 other co-travelers - this seems redundant. Reducing it may be a solution for harnessing exploding search space and keeping the efficiency of ride pooling.}}%
    \label{fig:AverageNode}%
\end{figure}

Finally we can see that in fact the average node degree in shareability graph increases with the demand and discount. In the case of Amsterdam, the shared discount offered by the service has a major impact on the shareability graph, as shown in Figure~\ref{fig:AverageNode}. The shareability graph remains robust for the mid-size demand level until the higher level of demand and discount up to 40\% as it may require a high computational cost to find the optimal number of trips until it reaches the critical point and explodes, the search becomes intractable.

\section{Conclusion}

This paper investigates the computational complexity of the ride-pooling problem. This allowed us to study the curse of dimensionality and may provide the foundation to develop novel efficient methods or heuristics.  Our evaluation results show that the complexity grows strongly with the sizes of the ride-pooling problem. Search space and calculation times grow faster than linearly.
Thanks to effective utility-driven heuristics, we can dramatically reduce this search space, yet it still explodes under specific settings. In particular, the complexity depends mainly on the discount offered. This has the greatest impact on the size of the search space. The complexity also grows with demand levels, yet this impact is not so pronounced. We observed that there is a strong correlation between the complexity of ride pooling and its efficiency, yet only to some extend.

However, our findings provide convincing evidence that shows the impact of $\lambda$ on running time and search space complexity, and for the mid-size demand of 500 trip request, the computation time increases to 100 seconds and for 15\% discount, the search space grows $10^{5}$ rides for the batch of 500 trips and when the discount increases to 40\% where the computation becomes intractable. The main driver of the search space explosion is in the rides of thirds degree and more, which reach up to 150 000 feasbile trips in our experiments, while number of pairs did not exceeded 25 000.

Notably, the efficiency of ride-pooling does not necessarily comes at the cost of enormous search space. After the critical mass is reached the ride-pooling efficiency plateaus while search space explodes. This suggest that for increasing sizes of practical problems the search space may be smartly filtered. We identify a strong relation between the topology of the shareability graph and a search space size. Smart controlling of the shareability graph topology is a promising direction to improve efficiency of the ride-pooling algorithms without negatively impacting the efficiency of pooling itself.

\section{Acknowledgements}
This research is funded by National Science Centre in
Poland program OPUS 19 (Grant Number 2020/37/B/HS4/01847).

\section{Contribution}
The authors confirm contribution to the paper as follows: \textbf{Usman Akhtar:} Conceptualization, Methodology, Investigation, Implementation, Writing - Original Draft, \textbf{Rafał Kucharski:} Conceptualization, Methodology, Writing - Review \& Editing, Supervision

\newpage

\bibliographystyle{trb}
\bibliography{trb_template}
\end{document}